\documentclass{WileyMSP-template}

\usepackage{amsmath}
\usepackage{amssymb}
\usepackage{physics}
\usepackage{upgreek}
\usepackage{bm}
\usepackage{cite}
\usepackage[colorlinks=true]{hyperref}
\usepackage{mathptmx}

\usepackage{ragged2e}
\AtBeginDocument{\justifying}  

\newcommand{\rev}[1]{{\leavevmode\color{black}#1}}

\begin{document}

\pagestyle{fancy}
\rhead{\includegraphics[width=2.5cm]{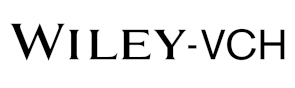}}

\title{Observation of dislocation bound states and skin effects in non-Hermitian Chern insulators}

\maketitle


\author{Jia-Xin Zhong}
\author{Bitan Roy*}
\author{Yun Jing*}



\begin{affiliations}
J.-X. Zhong, Y. Jing\\
{Graduate Program in Acoustics, The Pennsylvania State University, University Park, PA 16802, USA}\\
Email Address: yqj5201@psu.edu

B. Roy\\
{Department of Physics, Lehigh University, Bethlehem, Pennsylvania 18015, USA}\\
Email Address: bitan.roy@lehigh.edu

\end{affiliations}


\keywords{Non-Hermitian physics, topological acoustics, dislocation, defect}

\begin{abstract}

The confluence of non-Hermitian (NH) topology and crystal defects has culminated significant interest, yet its experimental exploration has been limited due to the challenges involved in design and measurements. Here, we showcase experimental observation of NH dislocation bound states (NHDS) and the dislocation-induced NH skin effect in two-dimensional acoustic NH Chern lattices. By embedding an edge dislocations-antidislocation pair in such acoustic lattices and implementing precision-controlled hopping and onsite gain/loss via active meta-atoms, we reveal robust defect-bound states localized at dislocation cores within the line gap of the complex energy spectrum. 
\rev{We experimentally identify the emergence of bulk exceptional points (EPs) via spectral coalescence and phase rigidity analysis. 
We demonstrate that the NHDS survive against moderate NH perturbations but gradually delocalize and merge with the bulk (skin) states driven by these EPs under periodic (open) boundary conditions. }
Furthermore, our experiments demonstrate that the dislocation core can feature weak NH skin effects when its direction is perpendicular to the Burgers vector in periodic systems. Our findings, therefore, pave an experimental pathway for probing NH topology via lattice defects and open new avenues for defect-engineered topological devices.

\end{abstract}


\section{Introduction}
Crystalline defects, such as dislocations~\cite{Ran2009OnedimensionalTopologicallyProtected, TeoKanePRB2010, AsahiNagaosaPRB2012, JuricicPRL2012, SlagerNatPhys2012, JuricicPRB2014, HughesPRB2014, Hamasaki2017, Nayak2019ResolvingTopologicalClassification, QueirozPRL2019, Roy2021DislocationBulkProbe, Nag2021CommunPhys, Yamada2022BoundStatesPartial, DasRoyPRB2023} and disclinations~\cite{TeoHughesPRL2013, Benalcazar2014ClassificationTwodimensionalTopological, Liu2021BulkDisclinationCorrespondence, Peterson2021TrappedFractionalCharges}, play pivotal roles in the identification of topological crystals in nature and topological metamaterials in laboratory~\cite{Lin2023TopologicalPhenomenaDefects, Li2018TopologicalLighttrappingDislocation, Wang2021VortexStatesAcoustic, Xue2021ObservationDislocationInducedTopological}. In Hermitian topological systems, these defects can host robust bound states, protected by a combination of crystalline symmetry and topological invariant as dictated by the bulk-defect correspondence, analogous to the celebrated bulk-boundary correspondence. 
The emergence of non-Hermitian (NH) physics, where gain/loss and nonreciprocity play significant roles, introduces radical departures from conventional Hermitian behavior~\cite{Lee2016AnomalousEdgeState, Leykam2017EdgeModesDegeneracies, Kunst2018PRL, Bergholtz2021ExceptionalTopologyNonHermitian, Kawabata2018AnomalousHelicalEdge, kawabata2019bPRX, ThomalePRB2019, FangPRL2020, Borgnia2020NonHermitianBoundaryModes, Xiao2022TopologyNonHermitianChern, Bartlett2023UnravellingEdgeSpectra}. For instance, NH systems often exhibit phenomena such as the non-Hermitian skin effect (NHSE), in which a macroscopic number of eigenstates accumulate at the boundaries, obscuring the traditional bulk-boundary signature~\cite{Yao2018EdgeStatesTopological, Zhang2021AcousticNonHermitianSkin, Bhargava2021NonHermitianSkinEffect, Schindler2021DislocationNonHermitianSkin, Panigrahi2022NonHermitianDislocationModes, MannaRoyCommPhys2023, Chadha2024RealspaceTopologicalLocalizer, Liu2024MeasurementChernNumber, KoziFuPRB2024}. These effects necessitate alternative strategies for detecting and characterizing NH topological phases.
\vspace{\baselineskip} %

Recent theoretical advances have identified crystalline defects as robust local topological probes in NH systems, potentially overcoming the limitations imposed by the NHSE~\cite{Panigrahi2022NonHermitianDislocationModes, Banerjee2024TopologicalDisclinationStatesa}. 
\rev{Fundamentally, such defect topology can be classified into point-gap \cite{Bhargava2021NonHermitianSkinEffect, Schindler2021DislocationNonHermitianSkin} and line-gap \cite{Panigrahi2022NonHermitianDislocationModes} phases. While point-gap topology is typically characterized solely by the NHSE, line-gap topology is particularly intriguing as it can host both the dislocation-induced NHSE (D-NHSE) and topological bound states, besides displaying conventional NHSE.}
In this context, Panigrahi \emph{et al.}\ theoretically showed that dislocations in a \rev{line-gap} NH Chern insulator can host robust in-gap bound states, protected by pseudo-particle-hole symmetry, whose spatial localization and stability are governed by both NH parameters and the defect geometry~\cite{Panigrahi2022NonHermitianDislocationModes}. 
Moreover, an interplay between the dislocation and NHSE can give rise to a D-NHSE, manifesting a typical skin accumulation around the defect core under periodic boundary conditions (PBCs)~\cite{Bhargava2021NonHermitianSkinEffect, Schindler2021DislocationNonHermitianSkin, Panigrahi2022NonHermitianDislocationModes}. However, experimental observation of these fascinating phenomena has remained far from reality due to the requirement for delicate defect engineering combined with precision-controlled NH perturbations, along with the inherent difficulty in probing complex energy spectrum~\cite{ Xiong2024TrackingIntrinsicNonHermitian, Gu2022TransientNonHermitianSkin, Gao2024ControllingAcousticNonHermitian, Jiang2024ObservationNonHermitianBoundarya, Huang2024ComplexFrequencyFingerprintB, Zhong2025ExperimentallyProbingNonHermitian}.
\vspace{\baselineskip} %

In this work, we circumvent these challenges by embedding a pair of edge dislocation and antidislocation into a two-dimensional (2D) acoustic lattice implemented using a coupled acoustic cavity system (CACS) and by introducing controlled NH perturbations via active meta-atoms~\cite{Zhang2021AcousticNonHermitianSkin, Liu2022ExperimentalRealizationWeyl, Zhang2023ObservationAcousticNonHermitian, Chen2024RobustTemporalAdiabatic, Zhang2024NonreciprocalAcousticDevices, Zhong2025HigherorderSkinEffect, Zhong2025ExperimentallyProbingNonHermitian}.
Furthermore, the active meta-atoms are sophistically tuned to induce imaginary hopping terms, thereby breaking the time-reversal symmetry, an essential requirement for realizaing a Chern insulator.
This approach offers a significantly greater flexibility compared to conventional flow-based methods~\cite{Yang2015TopologicalAcoustics,Ding2019ExperimentalDemonstrationAcoustic}.
By employing a recently proposed Green's function approach \cite{Zhong2025ExperimentallyProbingNonHermitian}, we directly measure the complex-valued energy spectrum along with both left and right eigenstates of the system. Our experimental platform thereby enables the {direct} observation of NH dislocation states (NHDS) emerging within the line gap of the complex energy spectrum, as well as the D-NHSE. We further demonstrate that while the NHDS persist under moderate NH perturbations, they lose support in the system with the appearance of the exceptional points (EPs) therein. 
\rev{To validate this mechanism, we explicitly observe the bulk EPs in a pristine lattice, characterized by spectral coalescence and phase rigidity collapse.}
As this threshold is approached from the weak NH perturbation side, NHDS gradually delocalize and merge into the bulk (skin) states under periodic (open) boundary conditions; a process coinciding with the closure of the NH band gap, in quantitatively agreement with theoretical predictions. Overall, our work bridges the gap between NH band topology and defect physics, establishing crystal defects as universal tools for probing NH topological phases of matter.
\rev{We emphasize that our realization of a line-gap NH Chern insulator enables the observation of the coexistence of D-NHSE and topological defect-bound states and their interplay with EPs, phenomena that are fundamentally distinct from the D-NHSE characteristic of point-gap systems \cite{Wu2025ObservationDislocationNonHermitian}.}
\vspace{\baselineskip} %




\section{Results and Discussion}

\subsection{Model and Design}
\begin{figure}[!htb]
    \centering
    \includegraphics[width=.49\linewidth]{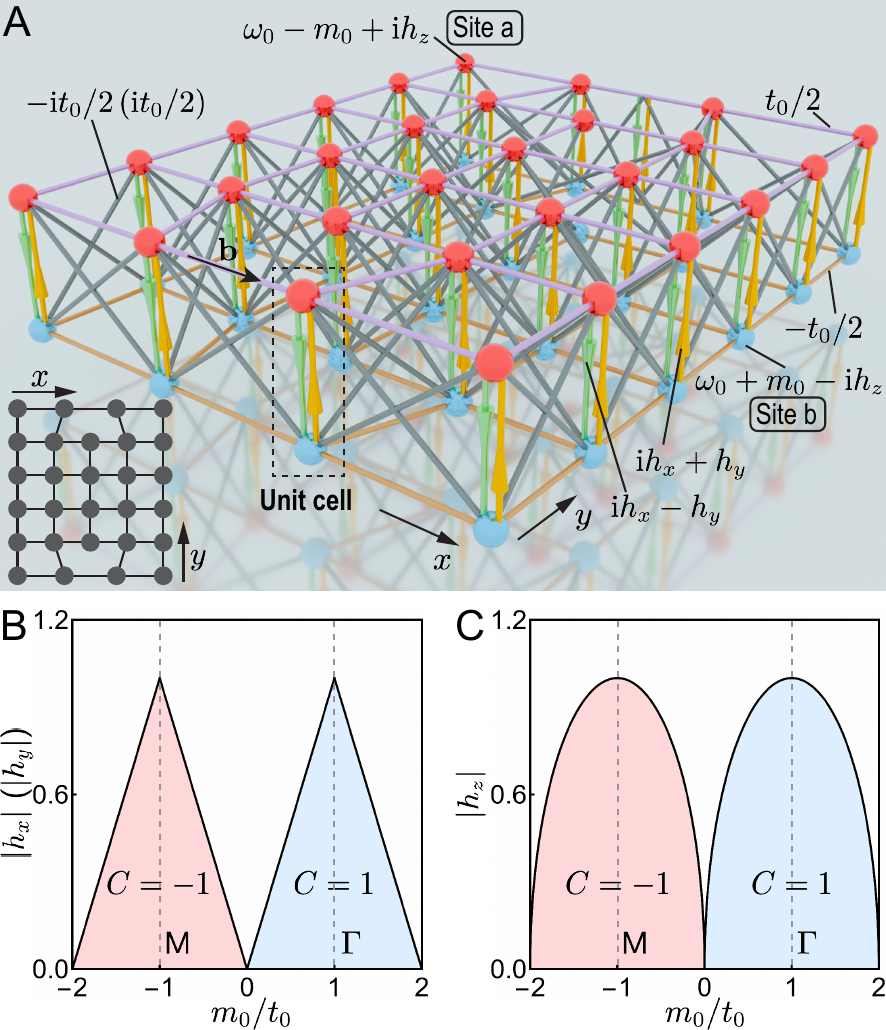}
    \caption{
        (A) Schematic implementation of a NH Chern insulator [Eq.~\eqref{eq:Hamiltonian}] in the presence of an edge dislocation-antidislocation pair with Burgers vectors $\vb{b} = \pm a \vb{e}_x$, where $\vb{e}_x$ is the unit vector in the $x$ direction and $a$ is the lattice spacing, set to be unity. Each unit cell consists of two sites labeled by a (red ball) and b (blue ball). Solid bars of the same color represent the reciprocal hopping amplitudes of equal strength, and those with arrows indicate nonreciprocal hopping amplitudes. Bottom left inset shows the top view, where circles represent unit cells.
        (B, C) Phase diagram of NH Chern insulators with NH perturbations for (B) $h_x$ (or $h_y$) and (C) $h_z$.
        The eigenenergies are line-gapped only in the red and blue shaded regions, where the NH Chern number is $C=-1$ and $C=1$, respectively. White regions support EPs. Experimental measurements are performed along the dashed gray lines.
    }
    \label{fig:1}
\end{figure}

We consider a 2D NH Chern insulator whose Bloch Hamiltonian is expressed as~\cite{Kawabata2018AnomalousHelicalEdge, Panigrahi2022NonHermitianDislocationModes, QWZPRB2006} 
\begin{equation}~\label{eq:Hamiltonian}
    H_\mathrm{NH}(\vb{k}) = H_\mathrm{H}(\vb{k}) + \mathrm{i} \bm\upsigma \vdot \vb{h}
    = \bm\upsigma \cdot \left[\vb{d}(\vb{k}) + \mathrm{i} \vb{h}\right],
\end{equation}
where $\vb{d}(\vb{k}) = \left( t_0 \sin (k_xa) , t_0\sin (k_ya) , t_0[ \cos (k_xa) + \cos (k_ya)] - m_0 \right)$, the vector Pauli matrix $\bm\sigma=(\sigma_x,\sigma_y,\sigma_z)$, $\mathrm{i}=\sqrt{-1}$ acts on two sublattices a and b, and we set the lattice spacing $a=1$. The Hermitian part is denoted as $H_\mathrm{H}(\vb{k}) = \bm\sigma\cdot \vb{d}(\vb{k})$ and the anti-Hermitian operator is $\mathrm{i} \bm\sigma\cdot \vb{h}$ with $\vb{h}=(h_x,h_y,h_z)\in \mathbb{R}^3$. 
In this model, an edge dislocation-antidislocation pair is introduced by removing two unit cells located at opposite ends along the $y$-direction and subsequently reconnecting the sites along the $x$-direction, see Fig.~\ref{fig:1}A. As a result, any closed path encircling the dislocation core lacks a translation by the Burgers vector $\vb{b} = \pm \vb{e}_x$.
The phase diagram of this model is shown in Figs.~\ref{fig:1}B and \ref{fig:1}C for nonzero $h_x$ or $h_y$ and $h_z$, respectively.
The NH Chern number is $C=-1$ in the red-shaded region and $C=1$ in the blue-shaded region, while it is trivial outside these areas. See Supplemental Information (SI) Appendix Sec.~S1 for details.
The system supports topologically protected dislocation states only in the red-shaded regions \cite{Panigrahi2022NonHermitianDislocationModes}.
In the Hermitian limit ($\vb{h}=0$) the above model features two distinct topological phases for $0<m_0/t_0<2$ and $-2<m_0/t_0<0$, respectively, featuring band inversion at the $\Gamma$ and M points of the Brillouin zone, therefore, named $\Gamma$ phase and M phase.
\vspace{\baselineskip} %

We implement this NH Chern insulator using a CACS, \rev{as illustrated in Fig.~\ref{fig:exp_setup_tuning}, which details the experimental setup and parameter tuning methodology}. 
See Experimental section along with SI Appendix Sec.~S2 for experimental details, including overall structure and tuning of onsite potentials and hoppings.
\rev{In our design, each acoustic cavity functions as a lattice site, configured to support a dipole resonant mode at the frequency $\omega_0 = 1040\,\mathrm{Hz} - 4.5\mathrm{i}\,\mathrm{Hz}$, where the imaginary component represents the intrinsic background loss. 
It is important to note that $\omega_0$ denotes the reference onsite potential engineered for the NH Chern lattice, rather than the intrinsic resonant frequencies of the as-fabricated cavities. 
In practice, the natural frequencies of the original cavities deviate significantly due to inevitable fabrication errors and installation imperfections (see SI Appendix Fig.~S5). 
This inherent disorder poses a major challenge to the observation of precise topological phenomena. To overcome this, we employ an active self-feedback mechanism (Fig.~\ref{fig:exp_setup_tuning}B) where the signal picked up by a microphone is fed back to a speaker within the same cavity to precisely calibrate and lock the resonance of all 56 cavities to the uniform reference $\omega_0$ (see SI Appendix Sec.~S2.2 for details).
In Fig.~\ref{fig:exp_setup_tuning}C1, the target onsite potential is $\omega_0-m_0+\mathrm{i}\,h_z=1043\,\mathrm{Hz}-4.2\,\mathrm{i\,Hz}$, whereas the value extracted by fitting the experimental data is $1042.98\,\mathrm{Hz}-4.21\,\mathrm{i\,Hz}$. 
In Fig.~\ref{fig:exp_setup_tuning}C2, the target onsite potential is $\omega_0+m_0-\mathrm{i}\,h_z=1037\,\mathrm{Hz}-4.8\,\mathrm{i\,Hz}$, with the fitted value of $1037.00\,\mathrm{Hz}-4.81\,\mathrm{i\,Hz}$. 
Here, $m_0=-3\,\mathrm{Hz}$ and $h_z=0.3\,\mathrm{Hz}$. 
These two representative examples demonstrate the high precision of our onsite-potential tuning.
Further statistical data regarding the distribution precision and long-term stability of the onsite potentials across all 56 cavities are presented in SI Appendix Sec.~S2.5.
}

As illustrated in Fig.~\ref{fig:1}A, each unit cell comprises two coupled cavities labeled by `a' and `b'. 
The intra-unit-cell hopping between these two cavities is rendered by the nonreciprocal NH perturbations, yielding asymmetric hopping amplitudes $\mathrm{i}h_x - h_y$ and $\mathrm{i}h_x + h_y$, while the inter-unit-cell hopping amplitudes are reciprocal. 
Both the nonreciprocal and reciprocal hoppings are realized via detector-source pairs (i.e., microphone-loudspeaker systems)~\cite{Zhong2025ExperimentallyProbingNonHermitian, Zhong2025HigherorderSkinEffect}.
For instance, \rev{as shown in Fig.~\ref{fig:exp_setup_tuning}D}, the hopping from site 1 to site 2 is achieved by detecting the acoustic pressure at site 1 and then driving the loudspeaker at site 2 accordingly. 
{Pure imaginary hopping terms ($\pm \mathrm{i} t_0/2$), required to break time-reversal symmetry for realizaing a Chern insulator, are similarly implemented via active components.}
The magnitude and phase of each microphone-loudspeaker pair are meticulously adjusted to realize the desired hopping parameters and onsite potentials.
\rev{
For configuration in Fig.~\ref{fig:exp_setup_tuning}E, the target hopping strength is $\kappa_0 = t_0/2 = 1.5\,\mathrm{Hz}$, and the retrieved value through fitting experimental results is $\kappa_0 = 1.51\mathrm{\,Hz} +0.01\mathrm{i\,Hz}$. 
For configuration Fig.~\ref{fig:exp_setup_tuning}F, the target hopping strength is $\kappa_0 = - h_y+\mathrm{i}h_x  = -0.3\,\mathrm{Hz} +0.3\mathrm{i\,Hz}$, and the retrieved value through fitting experimental results is $\kappa_0 = -0.30\mathrm{\,Hz} + 0.30\mathrm{i\,Hz}$. 
These two representative examples demonstrate the high precision of our hopping strength tuning.
Further statistical data regarding the distribution precision and long-term stability of the hopping strengths across all 56 cavities are presented in SI Appendix Sec.~S2.5.
}

\rev{
    We highlight that this fully programmable active coupling scheme offers distinct capabilities compared to passive acoustic interconnects or topolectrical circuit. 
    Unlike passive tubes typically restricted to reciprocal Hermitian couplings, our active meta-atoms enable the independent tuning of nonreciprocal terms as well as complex-valued hoppings that explicitly break time-reversal symmetry, which are essential for accessing the line-gap Chern phase. 
    Furthermore, regarding electric circuits, we note a fundamental distinction in physical realization. 
    While circuits are powerful tools for emulating the Hamiltonian matrix structure \cite{Lee2018TopolectricalCircuits}, our platform realizes the Hamiltonian governing the propagation of an actual physical wave field.
     Consequently, our observation of defect-core localization and NHSE serves as a direct demonstration of these phenomena in a wave medium, offering valuable insights transferable to other wave-based platforms such as photonics and mechanics.}

\subsection{Experimental Measurements}
Probing the wave dynamics in NH lattices poses unique challenges due to the presence of complex-valued eigenenergies, which manifest as complex poles in the Green's functions of the system. Therefore, conventional single-site excitation techniques, relying on the real or simple complex frequency excitations, cannot selectively target these poles~\cite{Gu2022TransientNonHermitianSkin, Xiong2024TrackingIntrinsicNonHermitian, Gao2024ControllingAcousticNonHermitian, Jiang2024ObservationNonHermitianBoundarya, Huang2024ComplexFrequencyFingerprintB, Zhong2025HigherorderSkinEffect, Zhong2025ExperimentallyProbingNonHermitian}. 
To overcome this limitation, we employ a Green's function-based measurement method~\cite{Zhong2025ExperimentallyProbingNonHermitian}.
In our experiments, a source (loudspeaker) is sequentially activated at every site, depicted in Fig.~\ref{fig:1}A, and the resulting frequency responses are simultaneously captured by an array of microphones.
This procedure yields the full Green's function $G(\omega)$ as a function of the excitation frequency $\omega$.
\rev{Representative raw spectral responses measured at distinct lattice sites are provided in SI Appendix Sec.~S3. We note that directly identifying the topological features (NHDS or D-NHSE) from these raw point-to-point spectra is challenging due to the intrinsic background loss, which obscures the specific complex poles. However, by analyzing the full measured Green's function matrix, the complex-valued eigenenergies and corresponding right and left eigenstates are accurately extracted.}
\rev{Importantly, this Green's-function-based approach is universally applicable to general linear wave systems, including those utilizing passive coupling mechanisms (e.g., acoustic tubes). 
To demonstrate this generality, we successfully applied this method to extract the complex spectrum and eigenstates of a minimal two-cavity system connected by passive tubes (see detailed validation in SI Appendix Sec.~S2.8).}

\subsection{Parameter space}
To ensure a comprehensive observation, we choose four distinct values of each NH perturbation in experiments; $h_y = h_z = 0$ and $h_x/t_0 = 0.3, 0.6, 0.9, 1.1$, with analogous values for $h_y$ and $h_z$. We also include the Hermitian case ($\vb{h} = 0$), resulting in a total of 13 configurations. For each configuration, experiments were performed under open boundary conditions (OBCs) and PBCs in two distinct phases; the ${\rm M}$ phase with $t_0 = -m_0 = 3\,\mathrm{Hz}$ and ${\bf K}= (\uppi, \uppi)$, and the $\Gamma$ phase with $t_0 = m_0 = 3\,\mathrm{Hz}$ and ${\bf K}= (0,0)$. Here, ${\bf K}$ is the band inversion momentum. In total, 52 cases were explored, with all experimentally measured complex-valued energy spectra, left and right eigenstates, and corresponding theoretical predictions detailed in the SI Appendix Sec.~S6 and Figs.~S20--S71.

\begin{figure}[!htb]
    \centering
    \includegraphics[width=0.99\linewidth]{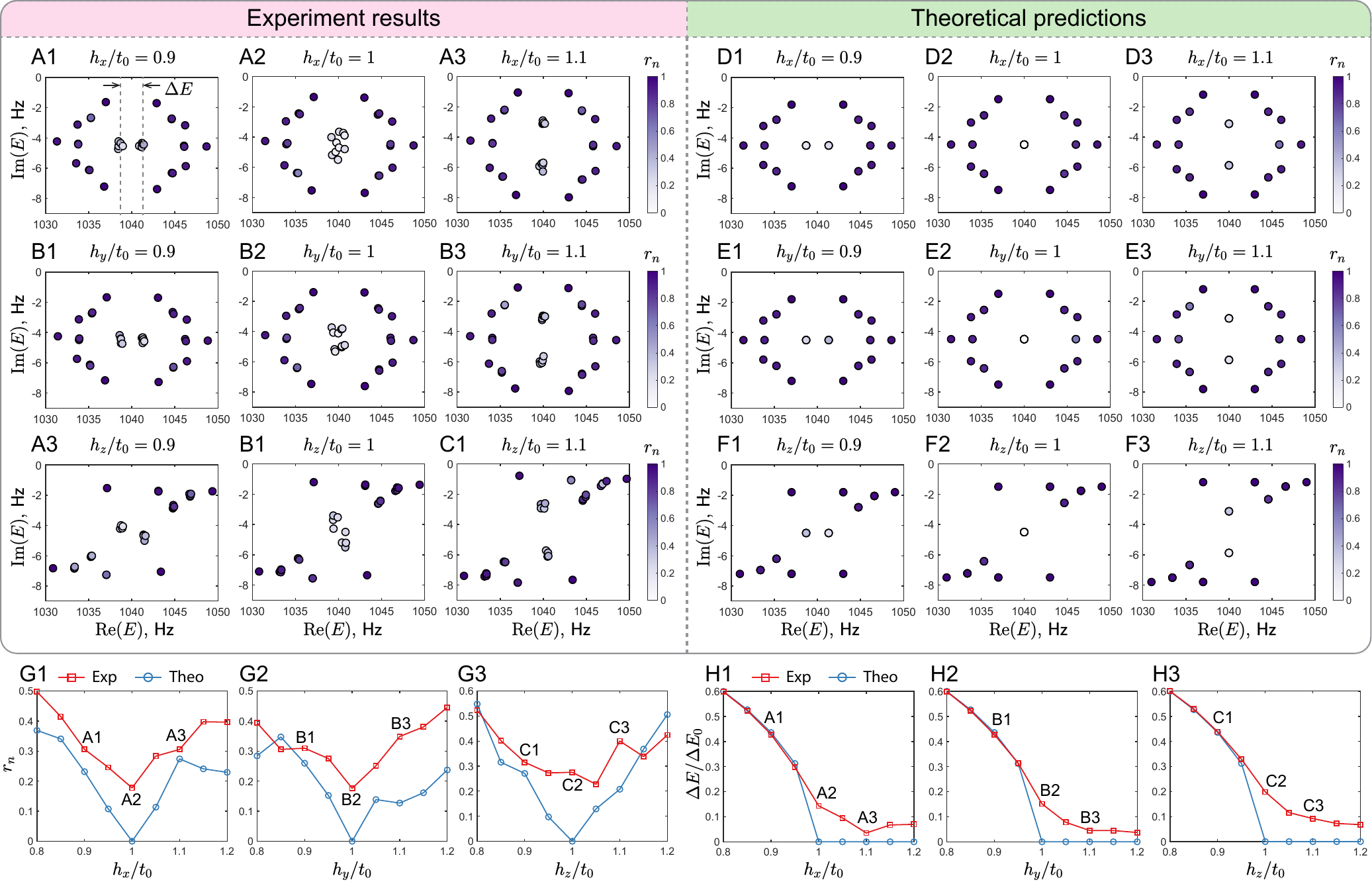}
    \caption{
        \rev{
        Experimental observation of bulk EPs in a pristine $4\times 4$ acoustic Chern lattice without dislocations under PBCs with NH perturbations for $t_0=-m_0 = 3\,\mathrm{Hz}$.
        (A, B, C) Experimental observations and (D, E, F) corresponding theoretical predictions of the complex energy spectra.
        The spectra are shown for varying NH perturbation strengths $h_x$ (A, D), $h_y$ (B, E), and $h_z$ (C, F) at values of $0.9t_0$ (left), $1.0t_0$ (middle), and $1.1t_0$ (right).
        The phase rigidity, $r_n$, of the $n$th state is indicated by the color bar.
        (G1--G3) Measured (red) and predicted (blue) averaged phase rigidity of all near-zero-energy states as a function NH perturbation strength $h_j$ for $j=x,y,z$.
        (H1--H3) The energy gap, $\Delta E$, defined in (A1) as a function of NH perturbation strength $h_j$ for $j=x,y,z$. Here, $\Delta E_0 = 6\,\mathrm{Hz}$ denotes the line gap in the Hermitian limit ($h_x=h_y=h_z=0$).
        }
    }
    \label{fig:bulk_ep}
\end{figure}

\rev{
\subsection{Observations of bulk EPs}
\label{sec:bulk_ep}

Before investigating the interplay between NH topology and dislocations, it is crucial to experimentally validate the emergence of EPs in the bulk lattice. We characterize the spectral properties of a pristine $4\times4$ lattice without dislocations under PBCs.
Figure~\ref{fig:bulk_ep} presents the experimental observation of EP signatures as the NH perturbations ($h_x, h_y, h_z$) are tuned across the theoretical critical value ($h_{j}/t_0=1, j=x,y,z$). Figures~\ref{fig:bulk_ep}A-F display the evolution of the complex energy spectra. 
The discrepancies between the experimental data and the theoretical predictions may arise from disorder in the onsite potentials and hopping amplitudes induced by experimental imperfections (see SI Appendix Sec.~S4 for details).
As the system approaches $h_\mathrm{c}$, we observe a distinct coalescence of eigenenergies in the complex plane, signaling the proximity to EPs.

To quantitatively verify the EP characteristics, we introduce the phase rigidity $r_n$ of the $n$th eigenstate, defined as
\begin{equation}
r_n = \frac{|\langle \psi_n^\mathrm{L} | \psi_n^\mathrm{R} \rangle|}{\sqrt{\langle \psi_n^\mathrm{L} | \psi_n^\mathrm{L} \rangle \langle \psi_n^\mathrm{R} | \psi_n^\mathrm{R} \rangle}},
\end{equation}
where $\ket{\psi_n^\mathrm{R}}$ and $\bra{\psi_n^\mathrm{L}}$ are the right and left eigenstates corresponding to the $n$th eigenenergy, respectively. 
The phase rigidity quantifies the biorthogonal overlap between left and right eigenstates. 
In a Hermitian system, $r_n=1$; however, near an EP, the eigenstates tend to coalesce, leading to a vanishing $r_n$. As shown in Figures~\ref{fig:bulk_ep}G1--G3, the averaged phase rigidity of all near-zero-energy states (crossed by dashed lines in Fig.~\ref{fig:bulk_ep}A1) exhibits a distinct minimum (dip) around $h/t_0=1$. This experimentally confirms that the system undergoes a bulk spectral phase transition driven by EPs.

Furthermore, Figures~\ref{fig:bulk_ep}H1-H3 reveal that the real-energy line gap ($\Delta E$, defined by the separation of two dashed lines in Fig.~\ref{fig:bulk_ep}A1) starts to close at this critical point. This bulk gap closure provides the physical mechanism for the ``melting'' of the NHDS observed in the subsequent sections. Since the topological protection of NHDS relies on the existence of an open line gap, its closure at the EP naturally leads to the delocalization of defect modes and their absorption into the bulk continuum.
}

\subsection{Observations of NHDS and D-NHSE}
\begin{figure*}[t]
    \centering
    \includegraphics[width=0.99\linewidth]{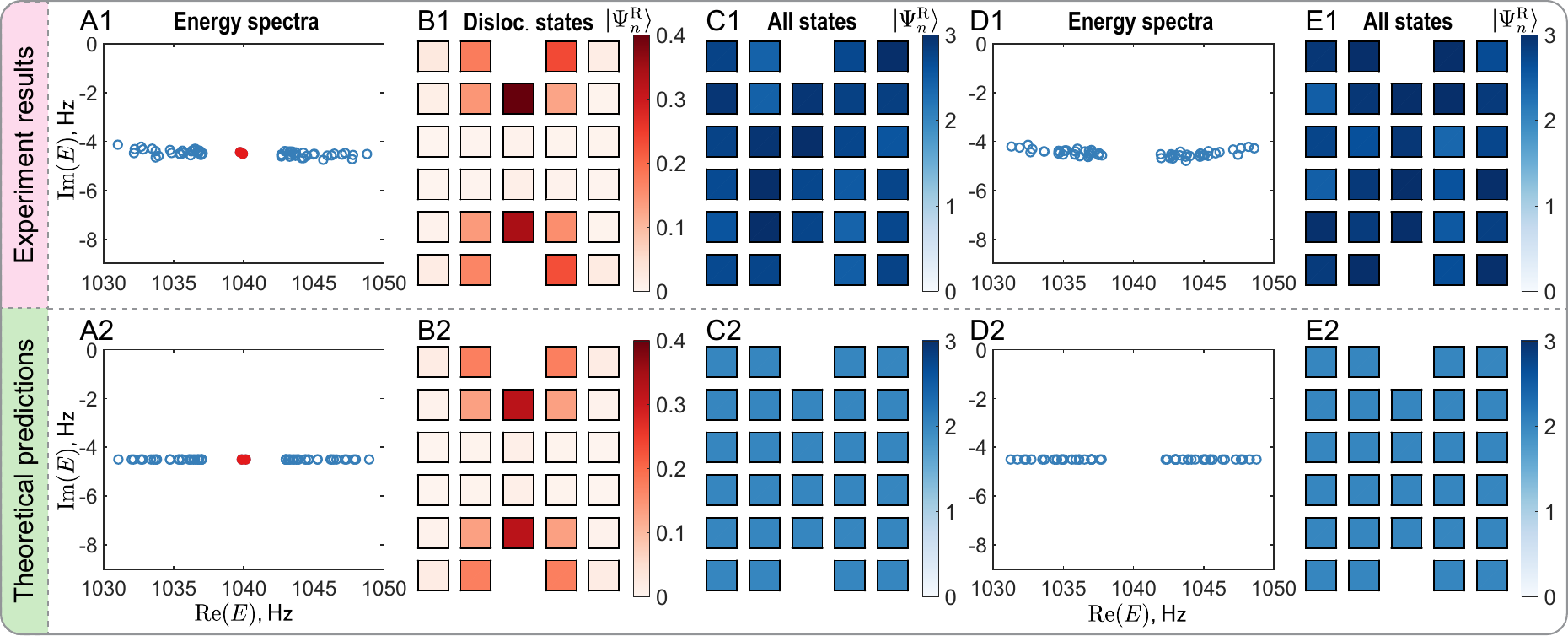}
    \caption{
        Hermitian ($\vb{h}=0$) acoustic Chern insulators in the presence of an edge dislocation-antidislocation pair under PBCs, showing (A1)-(E1) experimental observations and (A2)-(E2) theoretical computations. Energy spectrum [(A1) and (A2)], amplitude distributions of right eigenstates of the dislocation modes, shown by red dots in (A1) and (A2) within unit cells [(B1) and (B2)] and all the states [(C1) and (C2)] in the ${\rm M}$ phase with $t_0=-m_0 = 3\,\mathrm{Hz}$. Energy spectra [(D1) and (D2)] and amplitude distributions of all the right eigenstates [(E1) and (E2)] in the $\Gamma$ phase with $t_0=m_0 = 3\,\mathrm{Hz}$.
    }
    \label{fig:2}
\end{figure*}

We begin our investigation by examining the dislocation bound states in the Hermitian limit ($\vb{h} = \vb{0}$). Figure~\ref{fig:2} displays the experimental results for both the ${\rm M}$ and $\Gamma $ phases. The measured complex energy spectra and corresponding right eigenstates exhibit excellent agreement with theoretical predictions. Namely, only the M phase (Fig.~\ref{fig:2}A) hosts mid-gap dislocation states as ${\bf K}\cdot {\bf b}=\pm \pi$ (nontrivial) around the defect core therein, whereas the $\Gamma$ phase (Fig.~\ref{fig:2}D) is devoid of such defect modes for which ${\bf K}\cdot {\bf b}=0$ (trivial)~\cite{Ran2009OnedimensionalTopologicallyProtected, JuricicPRL2012, SlagerNatPhys2012}. The right eigenstate weight at the $n$-th unit cell is defined as 
$\Psi_n^\mathrm{R} = \sum_m \Bigl(\abs{\psi_{m,n,\mathrm{a}}^\mathrm{R}}^2 + \abs{\psi_{m,n,\mathrm{b}}^\mathrm{R}}^2\Bigr),$
where the index $m$ sums over either the two states near zero energy corresponding to the dislocation states (Fig.~\ref{fig:2}B) or all states (Figs.~\ref{fig:2}C and~\ref{fig:2}E). In this notation, $\psi_{m,n,\xi}^\mathrm{R}$ denotes the right eigenstate for the $m$th state at site $\xi$ within the $n$th unit cell, with $\xi={\rm a}, {\rm b}$. Notably, PBCs are imposed in Fig.~\ref{fig:2} to suppress midgap edge states, thereby enhancing the visibility of dislocation states. 
SI Appendix Figs.~S21 and S23 show corresponding results under OBCs.
\vspace{\baselineskip} %

As illustrated in Fig.~\ref{fig:2}A, two distinct midgap states are clearly observed only in the ${\rm M}$ phase, with their spatial profiles localized at the dislocation cores, see Fig.~\ref{fig:2}B. The dislocation modes are separated from the bulk states, see Fig.~\ref{fig:2}C. By contrast, the $\Gamma$ phase reveals only bulk eigenstates, with no midgap energies detected, see Figs.~\ref{fig:2}D1 and~\ref{fig:2}E1. These findings are consistent with the theoretical prediction that a Hermitian Chern insulator hosting a dislocation supports zero-energy states only in the ${\rm M}$ phase, where the first Chern number $C=-1$ \cite{JuricicPRL2012}, see also Figs.~\ref{fig:2}D2 and~\ref{fig:2}E2.

\vspace{\baselineskip} %

\begin{figure*}[!htb]
    \centering
    \includegraphics[width = 0.95\linewidth]{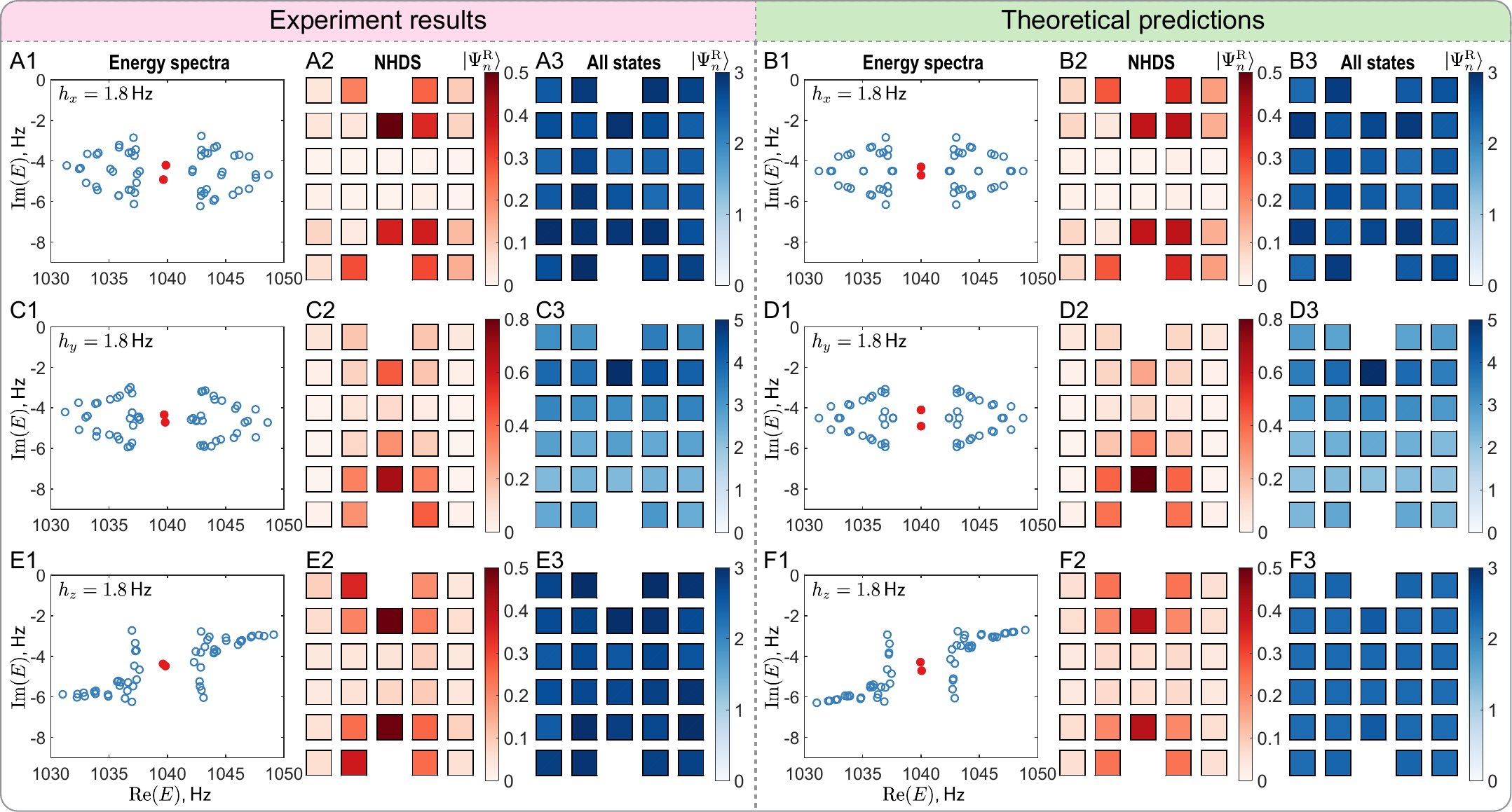}
    \caption{
        (A, C, E) Experimental observations and (B, D, F) corresponding theoretical predictions on an acoustic Chern lattice with an edge dislocation-antidislocation pair under PBCs and \emph{moderate} NH perturbations for $t_0=-m_0 = 3\,\mathrm{Hz}$. (A1, B1) Complex energy spectrum for $\vb{h}=(1.8\,\mathrm{Hz},0,0)$. (A2, B2) Total amplitude of two right eigenvectors of NHDS marked by red dots in (A1, B1). 
        (A3, B3) Total amplitude of all the right eigenstates from (A1, B1). Panels (C)-(D) [(E)-(F)] are analogous to (A)-(B), respectively, but with $\vb{h}=(0, 1.8\,\mathrm{Hz},0)$ [$\vb{h}=(0, 0, 1.8\,\mathrm{Hz})$]. 
    }
    \label{fig:3}
\end{figure*}

Upon introducing moderate NH perturbations, the energy spectra become considerably more intricate, as shown in Fig.~\ref{fig:3}. 
Nevertheless, two NHDS remain pinned near zero-energy within the line gap between 1038\,Hz and 1042\,Hz (approximately), distinguishing them from the point-gap associated with typical NHSE in other NH lattices~\cite{Yao2018EdgeStatesTopological}.
The spatial distribution of the corresponding eigenstates reveals their pronounced dependence on the NH perturbations in the system. Specifically, Figs.~\ref{fig:3}A2 and \ref{fig:3}B2 show that when only $h_x$ is nonzero, the weight of the right eigenstates is biased along the $x$-direction, with a greater concentration on the right side ($+x$) of the dislocation cores. This observation is consistent with the fact that this NH perturbation breaks (preserves) the reflection symmetry about $y$-axis ($x$-axis). 
Similarly, Figs.~\ref{fig:3}C2 and \ref{fig:3}D2 depict that for only a nonzero $h_y$, the eigenstate weight shifts along the $y$-direction, with an enhanced intensity near the bottom dislocation core, as this NH perturbation breaks (preserves) the reflection symmetry about $x$-axis ($y$-axis). By contrast, when only $h_z$ is nonzero, the intrinsic $C_4$ symmetry of $H_\mathrm{H}(\vb{k})$ is preserved, resulting in an approximately symmetric eigenstate distribution along both $x$ and $y$ directions, see Figs.~\ref{fig:3}E2 and \ref{fig:3}F2.

\vspace{\baselineskip} %

Furthermore, the geometry of dislocations (encoded in the Burgers vector) plays a crucial role in determining the fate of the D-NHSE under PBCs. 
As shown in Fig.~\ref{fig:1}A, the dislocation core possesses one (two) nearest-neighbor sites in the $y$ ($x$) direction when the Burgers vector is $\vb{b} = \pm \vb{e}_x$, thus mimicking the coordination number of a site living on an $y$ directional boundary. 
Consequently, under PBCs, no NHSE is observed for nonzero $h_x$ near the defect cores, see Figs.~\ref{fig:3}A3 and \ref{fig:3}B3 and SI Appendix Figs.~S24--S39, while nonzero $h_y$ yields a D-NHSE as shown in Figs.~\ref{fig:3}C3 and \ref{fig:3}D3. However, D-NHSE is relatively \emph{weak} compared to the conventional NHSE observed under OBCs at the exterior boundaries of the system. See SI Appendix Figs.~S40--S55 for a detailed comparison. Notably, when only $h_z$ is nonzero, no NHSE is observed anywhere in the system, irrespective of the boundary conditions, see Figs.~\ref{fig:3}E3 and \ref{fig:3}F3 and SI Appendix Figs.~S56--S71.

\subsection{Observations of the NHDS melting}
As the strength of NH perturbations increases, initially well-localized NHDS gradually spread away from the defect cores. Under PBCs, where the conventional NHSE at the exterior boundaries of the system is not supported, the weight of these states is progressively absorbed into the bulk. This trend is observed with different perturbation levels $h_j/t_0= 0.3, 0.6, 0.9$ for $j=x,y,z$, as detailed in SI Appendix Figs.~S24--S35, S40--S51, and S56--S67. A similar behavior is found for the left eigenstates. Nonetheless, in the presence of dislocations, the NH Chern insulators with $C=-1$ continues to support topologically robust localized states around the defect cores, which happens for $|h_j|<t_0$ when $|m_0|=t_0$.

\vspace{\baselineskip} %

\begin{figure*}[t!]
    \centering
    \includegraphics[width = 0.99\linewidth]{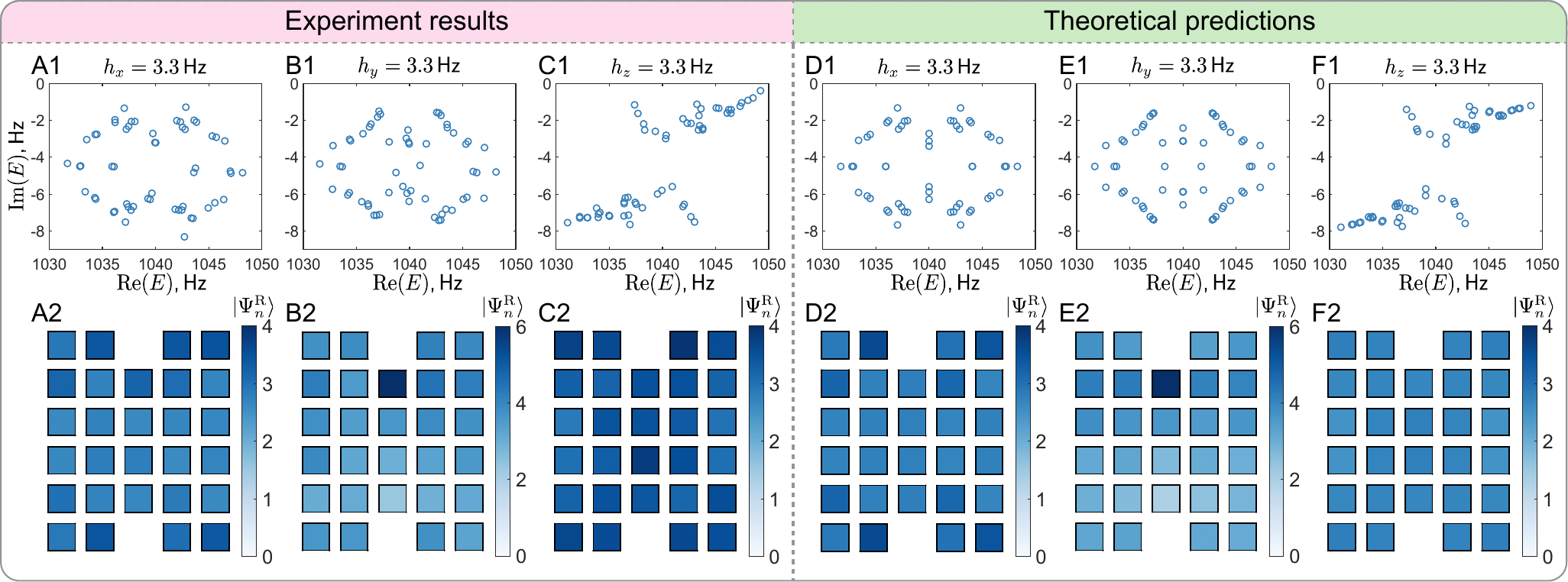}
    \caption{
        Similar to Fig.~\ref{fig:3}, but for \emph{strong} NH perturbations, showing the absence of NHDS. Complex energy spectrum for (A1, D1) $\vb{h}=(3.3\,\mathrm{Hz},0,0)$ (B1, E1) $\vb{h}=(0,3.3\,\mathrm{Hz},0)$, and (C1, F1) $\vb{h}=(0,0,3.3\,\mathrm{Hz})$. Corresponding total amplitude of all the right eigenstates are shown in (A2, D2), (B2, E2), and (C2, F2), respectively. 
        (A, B, C) Experimental observations and (D, E, F) corresponding theoretical predictions.
        }
    \label{fig:4}
\end{figure*}

For the parameter setting $t_0=-m_0=3\,\mathrm{Hz}$, the EPs first appear in the Brillouin zone when $h_j = 3\,\mathrm{Hz}$ for $j=x,y,z$, \rev{as validated in Fig.~\ref{fig:bulk_ep}}.
As the system approaches such a NH band gap closing, NHDS begin to lose their localization near the defect cores and gradually merge into the bulk states. This melting process is demonstrated in Fig.~\ref{fig:4}, where the system is placed slightly above the critical NH perturbation $h_j = 3.3\,\mathrm{Hz}$ for $j=x,y,z$, further detailed in SI Appendix Figs.~S39--S39, S52--S55, and S68--S71.
In the strong NH perturbation regime, the NHDS near zero-energy disappear, and only the summed weight of all right eigenstates is presented in Figs.~\ref{fig:4}A2--F2, showing that the localized defect states have fully merged into the bulk under PBC. 
As further corroborated by experiments under OBCs (see SI Appendix Figs.~S37, S53, and S69), the NHDS are found to completely merge into the skin states living at the exterior boundaries of the system. Collectively, Figs.~\ref{fig:3},~\ref{fig:4}, and the SI Appendix Figs.~S24--S71, provide clear evidence for a localization-delocalization phase transition in the NHDS, a phenomenon exclusively observable in NH systems.

\vspace{\baselineskip} %

It is noteworthy that Figs.~\ref{fig:4}B2 and \ref{fig:4}E2 illustrate a pronounced D-NHSE in the absence of NHDS for stronger $h_y$, with a higher right eigenstate weight at the top dislocation core compared to that for a weaker $h_y$, shown in Figs.~\ref{fig:3}C3 and \ref{fig:3}D3.
By contrast, for nonzero $h_x$ and $h_z$, the D-NHSE is either absent (likelihood) or much weaker, as shown in Figs.~\ref{fig:4}A2 and \ref{fig:4}C2, respectively. Therefore, our experimental observations clearly demonstrate a D-NHSE that is distinct from the conventional point-gap-supported NHSE under OBCs~\cite{Yao2018EdgeStatesTopological}, and it depends on the relative orientation of the Burgers vector and the directionality of the NHSE.   

\subsection{Observations under OBCs}
\begin{figure*}[t!]
    \centering
    \includegraphics[width = 0.7\linewidth]{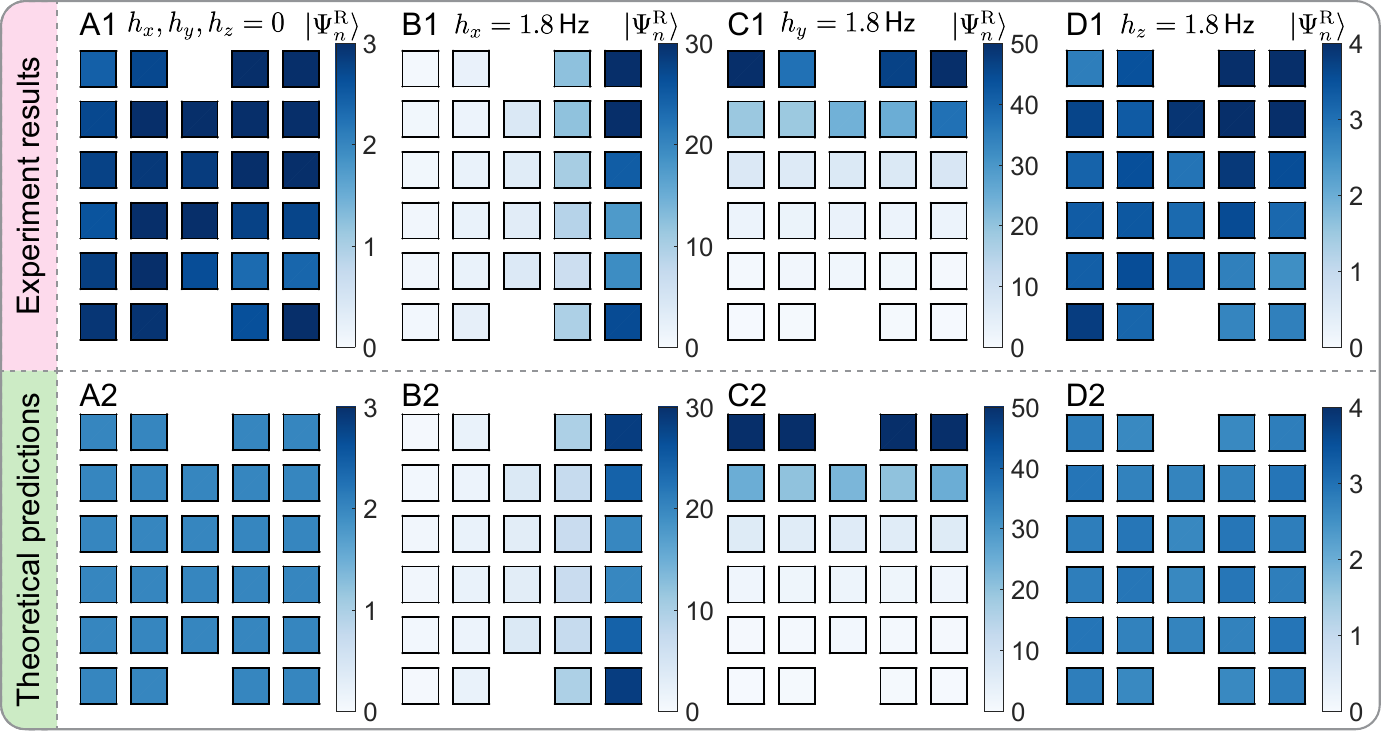}
    \caption{
        (A1--A4) Experimental observations and (B1--B4) theoretical predictions of the total amplitude of all right eigenstates under OBCs in the ${\rm M}$ phase with $t_0=-m_0=3\,\mathrm{Hz}$ for (A) $\vb{h}=(0,0,0)$ (Hermitian limit), (B) $\vb{h}=(1.8\,\mathrm{Hz},0,0)$, (C) $\vb{h}=(0, 1.8\,\mathrm{Hz},0)$, and (D) $\vb{h}=(0,0,1.8\,\mathrm{Hz})$.
        }
    \label{fig:5}
\end{figure*}

Under OBCs, the spatial distribution of the summed weight of all right eigenstates reveals valuable insights. 
For a Hermitian insulator and a NH one with nonzero $h_z$, such a weight extends uniformly throughout the entire lattice, as shown in Figs.~\ref{fig:5}A and~\ref{fig:5}D, respectively, confirming the absence of any NHSE therein. 
However, with the NH perturbations $h_x$ and $h_y$, clear signatures of NHSE emerge, with the right wavefunctions localizing toward the positive $x$ and $y$ boundaries, respectively, shown in Figs.~\ref{fig:5}B and~\ref{fig:5}C. This anisotropic behavior stems from the broken reflection and inversion symmetries by the NH perturbations $h_x$ and $h_y$. 
With the emergence of NHSE along the edges, the dislocation cores themselves do not exhibit a pronounced NHSE under OBCs for nonzero $h_x$ and $h_y$.
By contrast, when the Burgers vector is perpendicular to the direction of the NHSE observed under OBCs, a D-NHSE becomes visible under PBC, as show in Figs.~\ref{fig:4}B2 and \ref{fig:4}E2.

\section{Conclusion}
To summarize, our experiments provide conclusive evidences for NHDS and D-NHSE in a 2D acoustic Chern insulator. 
In the Hermitian limit, dislocations host zero-energy states in the ${\rm M}$ phase, consistent with the bulk-defect correspondence~\cite{Ran2009OnedimensionalTopologicallyProtected, JuricicPRL2012, SlagerNatPhys2012}. 
Under moderate NH perturbations, these states persist within the line gap.
As the perturbation strength increases, we observe a localization-delocalization transition for NHDS as they gradually spread and merge with bulk states under PBCs when the system approaches the shore of hosting EPs in the Brillouin zone~\cite{Panigrahi2022NonHermitianDislocationModes}. 
By contrast, the D-NHSE appears for any arbitrary strength of NH perturbations, when the direction of the conventional NHSE is orthogonal to the Burgers vector. 

\rev{Beyond their fundamental significance, these phenomena point toward specific applications in acoustic device engineering. For instance, the NHDS, which exhibits line-gap protected localization only within a specific spectral window, can serve as the basis for frequency-selective acoustic concentrators or filters. Conversely, the D-NHSE, which funnels bulk energy toward the dislocation core, suggests a mechanism for high-sensitivity sensing. The resulting intensity accumulation at the defect core can naturally amplify the system's response to minute local perturbations, such as mass loading or structural defects.}
Our results not only bridge the gap between NH band topology and defect physics but also establish crystalline defects as a universal tool for probing NH topological matter. 


\section{Experimental Section}

\begin{figure}[!htb]
    \centering
    \includegraphics[width=\textwidth]{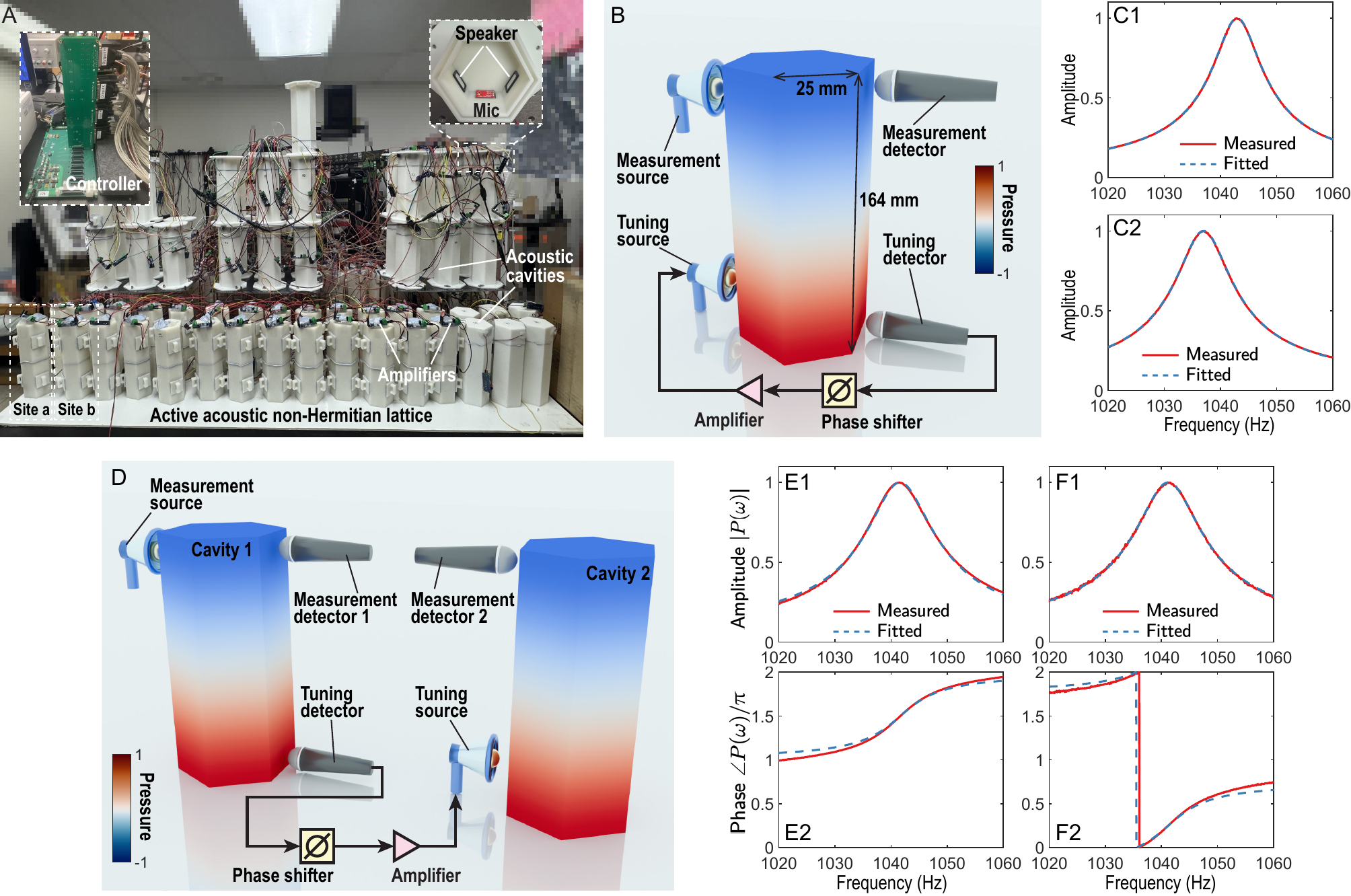}
    \caption{
        \rev{
    (A) A photograph of the experimental setup. The acoustic non-Hermitian lattice consists of an array of acoustic cavities, where 56 are used in the final configuration. 
    Onsite potential and hoppings are implemented using microphone-loudspeaker pair. 
    The hoppings are tuned using phase shifters integrated in a controller. 
    (B) Sketch of tuning onsite potential of acoustic cavities. 
    (C) Measured and fitted magnitude responses for two typical configurations that emulate the sublattices: (C1) sublattice `a' and (C2) sublattice `b' within a unit cell as annotated in Fig.~1A.
    (D) Schematic illustration of the unidirectional hopping implmeneted using a detector and a source.
    The bottom microphone (tuning detector) in cavity 1 captures the acoustic pressure signal, which is processed by the controller to adjust its phase and amplitude before being emitted by the bottom loudspeaker (tuning source) in cavity 2.
    Experimentally measured and numerically fitted amplitude (E1 and F1) and phase (E2 and F2) responses of the cross-power spectral density between the acoustic signals measured in cavities 1 and 2.
        }
    }
    \label{fig:exp_setup_tuning}
\end{figure}

\noindent\threesubsection{Experimental setup}\\
    Figure~\ref{fig:exp_setup_tuning}A shows a photograph of our experimental setup. 
    The system consists of 56 3D-printed acoustic cavities, each representing a site in the NH lattice. 
    A custom-designed digital controller is employed to precisely tune the onsite potentials of the acoustic cavities and the hoppings between them.
    The tuning of onsite potentials and hoppings is implemented using active components, specifically microphone-loudspeaker (detector-source) pairs. 
    The hopping implementations are independent of spatial distances between sites, enabling flexible implementation of various boundary conditions and lattice geometries with ease.
\vspace{\baselineskip} %

    The custom-made digital controller comprises three components: the core board, the motherboard, and the input/output (IO) board. 
    The core board houses a field-programmable gate array (FPGA, XC7K325T, Xilinx) and a digital signal processor (DSP, TMS320C6678, Texas Instruments), which enable real-time signal processing.
    During experiments, the system operates at a sampling frequency of 12.8\,kHz. 
    The motherboard supplies power and facilitates communication between the core and IO boards. 
    The IO board handles analog signal acquisition and output, interfacing with microphones and loudspeakers, respectively. 
    Each IO board supports 16 channels for both analog inputs and outputs. 
    The input channels utilize analog-to-digital converters (ADCs, ADC7606B, ADI), and the output channels employ digital-to-analog converters (DACs, DAC8568, Texas Instruments).
\vspace{\baselineskip} %
    
\noindent\threesubsection{Acoustic cavities and tuning of onsite potentials}\\
SI Appendix Sec.~S2.2 shows the tuning process of the onsite potentials of acoustic cavities.
The acoustic cavities used in this study were fabricated using 3D printing with a tolerance of 0.2\,mm or within 0.3\%. 
The material is LEDO 6060 photosensitive resin, which behaves as acoustically rigid for airborne sound. 
Each printed cavity has a wall thickness of 6\,mm. 
As shown in Fig.~\ref{fig:exp_setup_tuning}B, the cavities are hexagonal prisms with an interior height of $l = 164\,\mathrm{mm}$ and a side length of $25\,\mathrm{mm}$. 
To characterize the cavity in the experiments, a loudspeaker (source) excites the cavity, and a microphone (detector) measures the acoustic pressure.
The onsite potential, $\omega_0$, of a cavity is retrieved using the Green's function for a single site:
\begin{equation}
G_0(\omega) = -\frac{\Im(\omega_0)}{\omega - \omega_0},
\label{eq:green_single_site}
\end{equation}
where $\omega$ is the excitation frequency.
\vspace{\baselineskip} %

    \noindent\threesubsection{Tuning of hoppings}\\
    Figure~\ref{fig:exp_setup_tuning}D shows the tuning process of the hoppings.
    Both nonreciprocal and reciprocal hoppings between the cavities are implemented using detector-source (microphone-loudspeaker) pairs. 
    In our platform, the hopping strength and phase are precisely controlled by a digital multi-channel controller, allowing flexible and reconfigurable manipulation of lattice hoppings.
    Each unidirectional hopping is realized through a loudspeaker, a microphone, and an audio amplifier (Texas Instruments LM386). 
    The loudspeaker and microphone are positioned at the bottom of each cavity. 
    The microphone captures the acoustic pressure signal, which is processed by the controller to adjust its phase and amplitude before being emitted by the loudspeaker in the connected cavity. 
    The analytical solution of this model is described in SI Appendix Sec.~S2.4.
    \rev{Crucially, to avoid self-oscillation, we actively verify that the system remains in the stable regime by confirming that all eigenenergies in the reconstructed complex spectrum possess negative imaginary parts (see SI Appendix Sec.~2.6 for details).}
\vspace{\baselineskip} %

    The tuning of the hopping parameters is performed as follows.
    As illustrated in Fig.~\ref{fig:exp_setup_tuning}D, a microphone is placed inside cavity 1 to detect the acoustic pressure.
    The detected pressure signal is then phase-adjusted, amplified, and emitted by the loudspeaker positioned in cavity 2.
    To determine both the amplitude and the phase of the unidirectional hopping, we excite at cavity 1 using a loudspeaker (measurement source) and measure the acoustic pressure in both cavities using two microphones (measurement detector).
    The cross-power spectral density of these two measured pressure signals is expressed by
    \begin{equation}
        P(\omega)
        \equiv
        \frac{\psi_\mathrm{2}(\omega)}{\psi_\mathrm{1}(\omega)}
        =
        \frac{\kappa_0}{\omega-\omega_0},
        \label{eq:crs_power_ampl}
    \end{equation}
    where $\kappa_0$ is the tuned hopping.
\vspace{\baselineskip} %


    \noindent\threesubsection{Implementation of boundary conditions}\\
    \rev{Leveraging the fully programmable nature of our active coupling scheme, we can seamlessly switch between PBCs and OBCs without physical reconfiguration of the acoustic structure. Since all hoppings are implemented via microphone-loudspeaker pairs connected through digital circuits, PBCs are realized by electronically enabling the feedback loops that connect the boundary sites across the lattice edges. Conversely, OBCs are achieved simply by disabling these specific boundary coupling channels in the controller. A schematic illustration using a simplified one-dimensional chain model is provided in SI Appendix Sec.~S2.7 to clarify this implementation.}
\vspace{\baselineskip} %

    \noindent\threesubsection{Measurement process}\\
    To obtain the full Green's function matrix, an acoustic source (loudspeaker) is sequentially excited at each cavity site, and the acoustic pressure is measured in all cavities using microphones. 
    This procedure is systematically repeated for every site across the entire lattice.
    The complex-valued energy spectra as well as left and right eigenstates of NH acoustic Chern insulators are obtained based on the method proposed in Ref.~\cite{Zhong2025ExperimentallyProbingNonHermitian}.

\medskip
\textbf{Supporting Information} \par 
Supporting Information is available from the Wiley Online Library or from the author.

\medskip
\rev{
\textbf{Note added.} \par 
During the review process of this manuscript, we became aware of a related work by Wu \emph{et al.} \cite{Wu2025ObservationDislocationNonHermitian}. 
We note that the two works explore distinct and complementary regimes of non-Hermitian defect physics. 
Wu \emph{et al.} investigate the point-gap topology in a 2D Hatano-Nelson lattice, focusing on the D-NHSE. 
Conversely, our work realizes a line-gap non-Hermitian Chern insulator, which enables the observation of the coexistence of D-NHSE and topological defect-bound states (NHDS), as well as the interplay with EPs. Collectively, these studies establish a comprehensive experimental framework for understanding defect engineering across different non-Hermitian topological classes.
}

\medskip
\textbf{Acknowledgements} \par 
Y.J.\ thanks the support of startup funds from Pennsylvania State University and NSF awards 2039463, 195122, and 2401236. B.R.\ was supported by NSF CAREER Grant No.\ DMR-2238679. 

\medskip

%
\bibliographystyle{MSP}
\bibliography{bibtex.bib, bibtex2.bib}

\end{document}